\documentclass{ws-ijmpcs}

\begin{document}

\markboth{G. Oliveira-Neto \it et al.}
{An Early Universe Model with Stiff Matter and a Cosmological Constant}


\title{AN EARLY UNIVERSE MODEL WITH STIFF MATTER AND A COSMOLOGICAL CONSTANT}

\author{G. OLIVEIRA-NETO\footnote{permanent address}}

\address{Departamento de F\'{\i}sica, \\
Instituto de Ci\^{e}ncias Exatas, \\
Universidade Federal de Juiz de Fora,\\
CEP 36036-330 - Juiz de Fora, MG, Brazil\\
gilneto@fisica.ufjf.br}

\author{G. A. MONERAT, E. V. CORR\^{E}A SILVA, C. NEVES, L. G. FERREIRA
FILHO\footnote{permanent address of all four authors}}

\address{Departamento de Matem\'{a}tica, F\'{\i}sica e Computa\c{c}\~{a}o, \\
Faculdade de Tecnologia, \\
Universidade do Estado do Rio de Janeiro,\\
Rodovia Presidente Dutra, Km 298, P\'{o}lo
Industrial,\\
CEP 27537-000, Resende-RJ, Brazil.\\
monerat@uerj.br, evasquez@uerj.br, cliffordneves@uerj.br, gonzaga@uerj.br}

\maketitle

\begin{history}
\received{\today}
\revised{Day Month Year}
\end{history}

\begin{abstract}
In the present work, we study the quantum cosmology description of
a Friedmann-Robertson-Walker model in the presence of a stiff
matter perfect fluid and a negative cosmological constant. We work
in the Schutz's variational formalism and the spatial sections have
constant negative curvature. We quantize the model and
obtain the appropriate Wheeler-DeWitt equation. In this model the
states are bounded therefore we compute the discrete energy spectrum
and the corresponding eigenfunctions. In the present work, we consider
only the negative eigenvalues and their corresponding eigenfunctions. This
choice implies that the energy density of the perfect fluid is negative.
A stiff matter perfect fluid with this property produces a model with a 
bouncing solution, at the classical level, free from an initial 
singularity. After that, we use the
eigenfunctions in order to construct wave packets and evaluate the
time-dependent expectation value of the scale factor. We find that
it oscillates between maximum and minimum values. Since the
expectation value of the scale factor never vanishes, we confirm that
this model is free from an initial singularity, also, at the quantum level.

\keywords{Stiff matter; Wheeler-DeWitt equation; negative cosmological constant.}


\end{abstract}

\ccode{PACS numbers: 04.40.Nr,04.60.Ds,98.80.Qc}

\section{Introduction}
\label{sec:intro}

The great importance of cosmological models where the matter content is represented by a stiff matter
perfect fluid was recognized since its introduction by Zeldovich.\cite{zeldovich,zeldovich2} This perfect fluid
has an equation of state of the form, $p = \alpha w$, with $\alpha = 1$, where $w$ and $p$ are, respectively,
the fluid energy density and pressure. It can also be described by a massless free scalar field. In order
to understand better the importance of this perfect fluid for cosmology, one has to compute its energy
density. In the temporal gauge ($N(t) = 1$), this quantity is proportional to $1/a(t)^6$, where $N(t)$ is
the lapse function and $a(t)$ is the scale factor. On the other hand, in the same gauge, the energy
density of a radiative perfect fluid is proportional to $1/a(t)^4$. This result indicates that there
may have existed a phase earlier than that of radiation, in our Universe, which was dominated by
stiff matter. Due to that importance, many physicists have started to consider the implications of
the presence of a stiff matter perfect fluid in FRW cosmological models. The first important
implication of the presence of stiff matter in FRW cosmological models is in the relic abundance of
particle species produced after the `Big Bang' due to the expansion and cooling of our Universe.
\cite{turner,joyce,salati,pallis,gomez,pallis1} The presence of stiff matter in FRW cosmological
models may also help explaining the baryon asymmetry and the density perturbations of the
right amplitude for the large scale structure formation in our Universe.\cite{zeldovich1,joyce1}
It may also play an important role in the spectrum of relic gravity waves created during
inflation.\cite{sahni} Since there may have existed a phase earlier than that of radiation which
was dominated by stiff matter some physicists considered quantum cosmological models with this
kind of matter.\cite{nelson,nelson2,nelson1}

In the present work, we study the quantum cosmology description of
a Friedmann-Robertson-Walker (FRW) model in the presence of a stiff
matter perfect fluid and a negative cosmological constant. The model
has constant negatively curved spatial sections. We work in
the Schutz's variational formalism.\cite{schutz,schutz1} We quantize the model and
obtain the appropriate Wheeler-DeWitt equation. In this model the
states are bounded therefore we compute the discrete energy spectrum
and the corresponding eigenfunctions. In the present work, we consider
only the negative eigenvalues and their corresponding eigenfunctions. This
choice implies that the energy density of the perfect fluid is negative.
A stiff matter perfect fluid with this property has already been
considered in the literature.\cite{nelson3,finelli,nelson4} It produces 
a model with a bouncing solution, at the classical level, free from an 
initial singularity. After that, we use the
eigenfunctions in order to construct wave packets and evaluate the
time-dependent expectation value of the scale factor. We find that
it oscillates between maximum and minimum values. Since the
expectation value of the scale factor never vanishes, we confirm that
this model is free from an initial singularity, also, at the quantum level.

The presence of a negative
cosmological constant in the present model implies that the universe described
by it has a maximum size, in other words it is bounded. Taking into account the
current cosmological observations, a negative cosmological constant will not be
able to describe the present accelerated expansion of our Universe. It is not
our intention to describe the present state of our Universe with this model.
On the other hand, it is our intention to
describe a `possible' state of our primordial Universe. One important theory
which is a strong candidate to describe the unification of all known physical
interactions is superstring theory.\cite{green,green1} Due to that, many physicists believe that
superstring theory will correctly describe the quantum gravity effects that took place
at the beginning of our Universe. There is an important conjecture which tells
that Type IIB string theory on $(AdS_5 \times S_5)_N$ plus some appropriate boundary conditions
is dual to $N =4$ $d=3+1$ $U(N)$ super-Yang-Mills.\cite{maldacena} It means that, possibly,
for an appropriate description of the known physical interactions through superstring
theories, the strings have to exist in an Anti-DeSitter spacetime. The Anti-DeSitter
spacetime has a negative cosmological constant, therefore it seems worthwhile to study
spacetimes with a negative cosmological constant if one wants to understand more about a
`possible' initial state of our primordial Universe. Of course, after that initial state
the Universe would have to undergo a transition where the cosmological constant would
change sign. Besides that, several important theoretical results and predictions in quantum
cosmology have been obtained with a negative cosmological
constant.\cite{carlip,carlip1,gil2,gil3} Considering a subset of all four-dimensional
spacetimes with constant negative curvature and compact space-like hypersurfaces, S. Carlip et
al showed how to compute the sum over topologies leading to the {\it no-boundary}
wave-function.\cite{carlip,carlip1} These spacetimes are curved only due to the presence of a negative
cosmological constant. In Ref. \refcite{carlip} it was shown how to obtain a vanishing
cosmological constant as a prediction from the {\it no-boundary} wave-function and in Ref.
\refcite{carlip1} it was shown how to obtain predictions about the topology of the Universe
from the {\it no-boundary} wave-function. We may also mention the result in Ref. \refcite{gil2},
where the WKB {\it no-boundary} wave-function of a homogeneous and isotropic Universe with a
negative cosmological constant was computed. Due to the regularity condition imposed upon
the spacetimes contributing to the {\it no-boundary} wave-function, it was shown that only a
well defined, discrete spectrum for the cosmological constant is possible. It was also found
that among the spacetimes contributing to the wave function, there were two complex conjugate
ones that showed a new type of signature change.

The present paper is organized as follows.
In Sec. \ref{sec:classical}, we introduce the classical model and obtain the
appropriate Friedmann equation. With the aid of the potential curve
coming from this equation, we comment on the general behavior of the classical
solutions. In Sec. \ref{sec:quantum}, we quantize the model by solving
the corresponding Wheeler-DeWitt equation. The wave-function depends on
the scale factor $a$ and on the canonical variable associated to the fluid,
which in the Schutz variational formalism plays the role of time ($T$). We
separate the wave-function in two parts, one depending solely on the scale
factor and the other depending only on the time. The solution in the time
sector of the Wheeler-DeWitt equation is trivial, leading to an imaginary
exponential of the type $e^{-iE\tau}$, where $E$ is the system energy and $
\tau =- T$. The scale factor sector of the Wheeler-DeWitt equation gives
rise to an eigenvalue equation. We find approximate solutions. In Sec. \ref
{sec:results}, we construct wave packets from the eigenfunctions
and compute the time-dependent, expectation value of the scale
factor. We find that the expectation value of the scale
factor shows bounded oscillations. Since the expectation value of the scale
factor never vanishes, we confirm that this model is free from
a {\it big bang} singularity, also, at the quantum level. Finally, in
Sec. \ref{sec:conclusions}, we summarize the main points and results of
our paper.

\section{The Classical Model}
\label{sec:classical}

The present Friedmann-Robertson-Walker cosmological model is characterized by the
scale factor $a(t)$ and has the following line element,
\begin{equation}
\label{1}
ds^2 = - N^2(t) dt^2 + a^2(t)\left( \frac{dr^2}{1 + r^2} + r^2 d\Omega^2
\right)\, ,
\end{equation}
where $d\Omega^2$ is the line element of the two-dimensional sphere with
unitary radius, $N(t)$ is the lapse function and we are using the natural unit
system, where $\hbar=c=8\pi G=1$. In this model the spatial sections are some
closed three-dimensional solid with negative constant curvature, locally isometric
to $H^3$.\cite{thurston} The matter content of the model is represented by a
perfect fluid with four-velocity $U^\mu = \delta^{\mu}_0$ in the comoving coordinate
system used, plus a negative cosmological constant ($\Lambda$). The total
energy-momentum tensor is given by,
\begin{equation}
T_{\mu \nu} = (w+p)U_{\mu}U_{\nu} - p g_{\mu \nu} - \Lambda g_{\mu \nu}\, ,
\label{2}
\end{equation}
As mentioned above, here, we assume that $p = w$, which is the equation of
state for stiff matter.

Einstein's equations for the metric (\ref{1}) and the energy momentum
tensor (\ref{2}) are equivalent to the Hamilton's equations generated by
the total Hamiltonian $N(t){\cal H}$, where,
\begin{equation}
\label{3}
{\cal H} = -\frac{{p_a}^2}{12a} + 3a + \Lambda a^{3}+ \frac{p_T}{a^3}.
\end{equation}
The variables $p_a$ and $p_T$ are the momenta canonically conjugated to
the variables $a$ and $T$, respectively.

The classical dynamics is governed by the Hamilton's equations, derived from
the total Hamiltonian $N(t){\mathcal{H}}$ Eq. (\ref{3}). In the gauge
$N(t)=a(t)$, they are,

\begin{equation}
\left\{
\begin{array}{llllll}
\dot{a} = & \frac{\displaystyle \partial N\mathcal{H}}{\displaystyle
\partial p_{a}}=-\frac{\displaystyle p_{a}}{\displaystyle 6}\, , &  &  &  &
\\
&  &  &  &  &  \\
\dot{p_{a}} = & -\frac{\displaystyle \partial N\mathcal{H}}{\displaystyle
\partial a}=-6a - 4\Lambda a^3 + 2\frac{\displaystyle p_T}{\displaystyle a^3}
\, , &  &  &  &  \\
&  &  &  &  &  \\
\dot{T} = & \frac{\displaystyle \partial N\mathcal{H}}{\displaystyle
\partial p_{T}} = \frac{\displaystyle 1}{\displaystyle a^2}\, , &  &  &  &  \\
&  &  &  &  &  \\
\dot{p_{T}} = & -\frac{\displaystyle \partial N\mathcal{H}}{\displaystyle
\partial T}=0\, . &  &  &  &  \\
&  &  &  &  &
\end{array}
\right.
\label{4}
\end{equation}
Where the dot means derivative with respect to the conformal time $\tau \equiv
Nt$, which in the present gauge is equal to $at$.

We also have the constraint equation $\mathcal{H} = 0$. It gives rise to the
Friedmann equation,
\begin{equation}
\label{5}
\dot{a}^2 + V_c (a) = 0,
\end{equation}
where the potential $V_c (a)$ is equal to,
\begin{equation}
\label{6}
V_c (a) = -a^2 -\frac{\Lambda a^4}{3} - \frac{p_T}{3a^2}.
\end{equation}

\begin{figure}[pb]
\centerline{\psfig{file=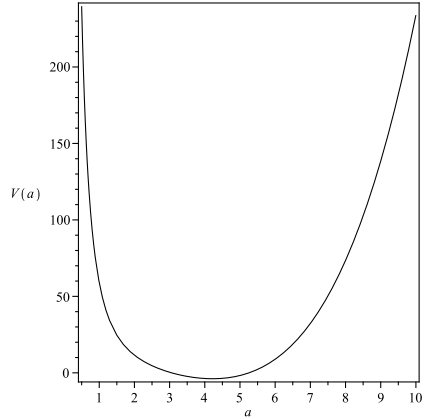,width=5.7cm}}
\vspace*{8pt}
\caption{$V_c (a)$ for $\Lambda = -0.1$ and $p_T = -180$. \label{figure1}}
\end{figure}

For the present situation where $\Lambda < 0$, we have bounded solutions. The
classical solutions are bouncing ones, free from an initial singularity, because 
we are considering a stiff matter perfect fluid with negative energy density.
Those results can be directly seen from the potential expression. A particular
example of $V_c (a)$, for $\Lambda = -0.1$ and $p_T = -180$, is given in
Figure \ref{figure1}.

\section{The Quantum Model}
\label{sec:quantum}

We wish to quantize the model following the Dirac formalism for quantizing
constrained systems.\cite{dirac,dirac1,dirac2,dirac3} First we introduce a
wave-function which is a function of the canonical variables $\hat{a}$ and
$\hat{T}$,

\begin{equation}
\label{7}
\Psi\, =\, \Psi(\hat{a} ,\hat{T} )\, .
\end{equation}
Then, we impose the appropriate commutators between the operators $\hat{a}$
and $\hat{T}$ and their conjugate momenta $\hat{p}_a$ and $\hat{p}_T$.
Working in the Schr\"{o}dinger picture, the operators $\hat{a}$ and $\hat{T}$
are simply multiplication operators, while their conjugate momenta are
represented by the differential operators
\begin{equation}
\hat{p}_{a}\rightarrow -i\frac{\partial}{\partial a}\hspace{0.2cm},\hspace{0.2cm}
\hspace{0.2cm}\hat{p}_{T}\rightarrow -i\frac{\partial}{\partial T}\hspace{0.2cm}.
\label{8}
\end{equation}

Finally, we demand that $\hat{{\cal H}}$, the superhamiltonian operator corresponding
to (\ref{3}), annihilate the wave-function $\Psi$ Eq. (\ref{7}), which leads to
Wheeler-DeWitt equation
\begin{equation}
\bigg(-\frac{1}{12}\frac{{\partial}^2}{\partial a^2} - ( 3a^2 +
\Lambda a^4 )\bigg)\Psi(a,\tau) = i \,\frac{1}{a^2} \frac{\partial}{\partial \tau}
\Psi(a,\tau),
\label{9}
\end{equation}
where the new variable $\tau= -T$ has been introduced.

The operator $\hat{{\cal H}}$ is self-adjoint \cite{lemos} with respect to the
internal product,

\begin{equation}
(\Psi ,\Phi ) =  \int_0^{\infty} da\, \frac{1}{a^2}\,\Psi(a,\tau)^*\, \Phi (a,\tau)\, ,
\label{10}
\end{equation}
if the wave-functions are restricted to the set of those satisfying either
$\Psi (0,\tau )=0$ or $\Psi^{\prime}(0, \tau)=0$, where the prime $\prime$
means the partial derivative with respect to $a$.

The Wheeler-DeWitt equation (\ref{9}) may be solved by writing $\Psi(a, \tau)$ as,
\begin{equation}
\Psi (a,\tau) = e^{-iE\tau}\,\eta(a)\,
\label{11}
\end{equation}
where $\eta(a)$ depends solely on $a$. Then $\eta(a)$ satisfies the eigenvalue equation
\begin{equation}
-\frac{d^2{\eta(a)}}{da^2} + V (a)\,\eta(a)= 12\,E\,\frac{1}{a^2}\,\eta(a)\, ,
\label{12}
\end{equation}
where the potential $V(a)$ is given by
\begin{equation}
V(a) = -36a^2-12\Lambda a^4\, .
\label{13}
\end{equation}

In the same way as in the classical regime, the potential $V(a)$ Eq. (\ref{13})
gives rise to bound states. Therefore, the energies $E$, Eq. (\ref{12}), of those
states will form a discrete set of eigenvalues $E_n$, where $n=1,2,3,...$. For each
eigenvalue $E_n$ there will be a corresponding eigenvector $\eta_n(a)$. The general
solution to the Wheeler-DeWitt equation (\ref{9}) is a linear combination of all
those eigenvectors,
\begin{equation}
\label{14}
\Psi(a,\tau) = \sum_{n=1}^{\infty} C_n\eta_n(a)e^{-iE_n\tau},
\end{equation}
where $C_n$ are free coefficients to be specified.

We are going to use the Galerkin or spectral method (SM), in order to solve the
eigenvalue equation (\ref{12}). This method is well presented in Ref. \refcite{boyd}
and it has already been used in quantum cosmology.\cite{pedram,pedram1} In the SM,
one must choose orthonormal basis functions in order to expand the solution to the
eigenvalue equation. The solutions to the present eigenvalue equation (\ref{12})
must fall sufficiently fast for large scale factor values ($a$). It means that we
must restrict the initial infinity domain of our variable $a$, to a finite domain.
Say, $0 < a < L$, where $L$ is a finite number that has to be fixed. Here, we
shall consider solutions satisfying the condition $\Psi (0,\tau )=0$. Putting
together all the above properties of the solutions to Eq. (\ref{12}), it is convenient
(but not mandatory) to choose our basis functions to be sine functions. Therefore,
we may write $\eta_n(a)$ in Eq. (\ref{12}) as,
\begin{equation}
\label{15}
\eta_n(a) = \sum_{n=1}^{\infty} A_n \sqrt{\frac{2}{L}}\sin
\left({\frac{n\pi a}{L}}\right),
\end{equation}
where the $A_n$'s will be determined by the SM. In the same $a$ domain, we may also
expand, in the same basis, the other two important functions of $a$ appearing in
Eq. (\ref{12}),
\begin{equation}
\label{16}
V(a)\eta_n(a) = \sum_{n=1}^{\infty} B_n \sqrt{\frac{2}{L}}\sin
\left({\frac{n\pi a}{L}}\right),
\end{equation}
\begin{equation}
\label{17}
\left(\frac{12}{a^2}\right)\eta_n(a) = \sum_{n=1}^{\infty} B_n' \sqrt{\frac{2}{L}}\sin
\left({\frac{n\pi a}{L}}\right),
\end{equation}
Where $V(a)$ is given in Eq. (\ref{13}) and the coefficients $B_n$ and $B_n'$ can be
easily determined. They are determined with the aid of Eq. (\ref{15}) and the fact
that the basis functions are orthonormal. The coefficients $B_n$ and $B_n'$
are given by,
\begin{equation}
\label{18}
B_n = \sum_{m=1}^{\infty} C_{m,n} A_m,
\end{equation}
\begin{equation}
\label{19}
B_n' = \sum_{m=1}^{\infty} C_{m,n}' A_m,
\end{equation}
where,
\begin{equation}
\label{20}
C_{m,n} = \frac{2}{L} \int_0^L \sin \left(\frac{m\pi a}{L}\right) V(a)
\sin \left(\frac{n\pi a}{L}\right) da,
\end{equation}
\begin{equation}
\label{21}
C_{m,n}' = \frac{2}{L} \int_0^L \sin \left(\frac{m\pi a}{L}\right)
\left(\frac{12}{a^2}\right)
\sin \left(\frac{n\pi a}{L}\right) da.
\end{equation}
Introducing the results Eqs. (\ref{15})-(\ref{21}) in the eigenvalue equation
(\ref{12}) and using the fact that the basis functions are orthonormal, we obtain,
\begin{equation}
\label{22}
\left(\frac{n\pi}{L}\right)^2 A_n + \sum_{m=1}^{\infty} C_{m,n} A_m =
E \sum_{m=1}^{\infty} C_{m,n}' A_m.
\end{equation}
In order to derive some numerical results we must fix a finite number of basis
functions, in other words, a finite number for the maximum value of the summation
indices. Let us call this number $N$. The greater the value of $N$, the closer
our results will be to the exact ones. We shall be restricted by our computational
resources. Equation (\ref{22}), may be written in a compact notation as,
\begin{equation}
\label{23}
D'^{-1}\, D\, A\, = E\, A\, ,
\end{equation}
where $D$ and $D'$ are $N \times N$ square matrices and their elements are obtained
from Eq. (\ref{22}). The solution to Eq. (\ref{23}) gives the eigenvalues and
corresponding eigenfunctions to the bound states of our quantum cosmology model.

It is important to mention that the most correct form of $\eta(a)$ in the limit when
$a \to 0$ is not given by Eq. (\ref{15}). In order to obtain that expression for $\eta(a)$,
one has to introduce the ansatz $\eta(a) = Ca^\alpha$ (where $C$ is a constant and $\alpha$
is a number to be determined) in Eq. (\ref{12}). After that, one has to discard the terms which
have as coefficients the cosmological constant ($\Lambda$) and the curvature of the spatial
sections ($k = -1$). The main motivation to discard those terms is that they are proportional
to $a^{\alpha+4}$ and $a^{\alpha+2}$ and should be less important, in the limit $a \to 0$, than 
the term whose coefficient is an algebraic equation for $\alpha$. Finally, one solves the 
resulting equation imposing that the coefficient of the only remaining term, proportional to 
$a^{\alpha-2}$, vanishes. This gives rise to a second order algebraic equation for $\alpha$, 
which solution satisfying the boundary condition $\eta(0) = 0$ is given by: $\alpha = 0.5 + 
\sqrt{0.25 + 12|E|}$. Therefore, the most correct form for $\eta(a)$ in the limit when
$a \to 0$ is given by,
\begin{equation}
\label{23,5}
\eta(a) = Ca^{0.5 + \sqrt{0.25 + 12|E|}}.
\end{equation}
We notice that it cannot represent the correct expression for $\eta(a)$ in the limit 
$a \to \infty$ because it diverges in that limit. That solution is already known in the 
literature.\cite{germano} 

\section{Energy Spectrum, Wave Packet and Mean Value.}
\label{sec:results}

In this section we will solve Eq. (\ref{12}) using the SM. In order to choose the
number of basis functions $N$ and the values of $\Lambda$, we performed the following
numerical procedures. First of all, in order to choose the values of $\Lambda$, we
solved numerically, the eigenvalue equation (\ref{23}) for several different values
of $\Lambda$ and fixed values of $N$ and $L$. We noticed that, although, we are free
to choose any value of $\Lambda$, the results accuracy for small absolute values of
$\Lambda$ is better than for large absolute values of $\Lambda$, for a given number of
basis functions $N$. This means that, if we use a large absolute value of $\Lambda$,
we need to increase $N$ to obtain the same accuracy of the case with a small absolute
value of $\Lambda$. This, of course, would increase the computation time. Therefore,
we have decided to use small absolute values of $\Lambda$. Taking in account this
considerations, we choose $\Lambda = -0.1$ in the rest of our paper. In order
to choose the value of $N$, we solved numerically, the eigenvalue equation (\ref{23})
for several different values of $N$ and fixed values of $\Lambda$ and $L$. Then, we
compared the eigenvalues coming from the different choices of $N$. We noticed that
only some of the first eigenvalues, for each different choice of $N$, remained the
same up to a satisfactory accuracy. Therefore, we have decided to use $N=100$ and take
only the first eighteen eigenvalues to construct the wave packet. The accuracy of
the eigenvalues, in this case, is in the tenth digit after the dot. Finally, it
is important to mention that based on comparisons with results of other models studied
with the SM,\cite{pedram1} we chose $L=6$, in the present case.

Now, using all these values: $N=100$, $\Lambda = -0.1$ and $L=6$, in the determinant
constructed from Eq. (\ref{23}), we solve it and obtain the first $100$ energy eigenvalues
for the present case. From those, we take the first $18$ energy levels and list them in
Table \ref{tableenergy}.

\begin{table}[ph]
\tbl{The eighteen lowest energy levels for a FRW model with $k=-1$,
$\Lambda = -0.1$, a stiff matter perfect fluid ($p=\rho$), $N=100$ and $L = 6$.}
{\begin{tabular}{|c|c|c|}
\hline
$E_1$ = -380.2201284331828 & $E_2$ = -342.1147751350869 & $E_3$ = -305.9147225014253  \\ \hline
$E_4$ = -271.6319016521779 & $E_5$ = -239.2791871064332 & $E_6$ = -208.8705235210961  \\ \hline
$E_7$ = -180.4210774184946 & $E_8$ = -153.9474207233139 & $E_9$ = -129.4677552918257  \\ \hline
$E_{10}$ = -107.0021912238143 & $E_{11}$ = -86.57309703051190 & $E_{12}$ = -68.20554802882278 \\ \hline
$E_{13}$ = -51.92791264850499 & $E_{14}$ = -37.77263895532904 & $E_{15}$ = -25.77734426711615 \\ \hline
$E_{16}$ = -15.98638945709977 & $E_{17}$ = -8.453288958554614 & $E_{18}$ = -3.244733126937446 \\ \hline
\end{tabular}
\label{tableenergy}}
\end{table}

\begin{figure}[pb]
\centerline{\psfig{file=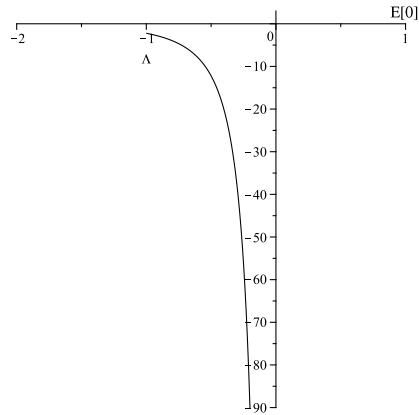,width=5.7cm}}
\vspace*{8pt}
\caption{Dependence of $E_1$ with $\Lambda$, for $N=100$ and $L = 6$.
\label{f4}}
\end{figure}

In order to give an idea how the energy spectrum depends on $\Lambda$,
we have constructed the curve of the fundamental energy  level $E_1$
versus $\Lambda$. It is given in Figure \ref{f4}. We notice that
$E_1$ decreases when one increases $\Lambda$.

It is important to mention that even though the expression of $\eta(a)$
given by Eq. (\ref{15}) is not the most correct one, in the limit $a \to 0$,
we were able to determine numerically that it converges rapidly to zero
in that limit. In particular, for the first energy level the corresponding 
eigenfunction oscillations about the zero value, in a region very close to 
it ($0 \leq a \leq 0.2$), are of the order of $10^{-12}$. For the eighteenth 
energy level the corresponding eigenfunction oscillations about the zero value, 
in a region very close to it ($0 \leq a \leq 0.04$), are of the order of 
$10^{-9}$. For the other eigenfunctions the oscillations about zero, in regions
very close to zero, have values between the ones given above.

\begin{figure}[pb]
\centerline{\psfig{file=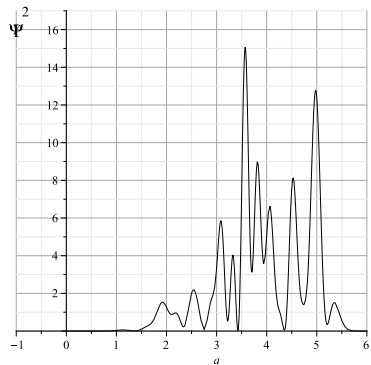,width=5.0cm}}
\vspace*{28pt}
\caption{Modulus squared of a wave packet constructed with the first
eighteen energy levels for $\tau = 1000$, $L = 6$ and $\Lambda$ = -0.1.
\label{f1}}
\end{figure}

Next, we construct the wave packet $\Psi (a, \tau )$ Eq. (\ref{14}),
with the aid of $\eta_n (a)$ Eq. (\ref{15}) and the energy levels in
Table \ref{tableenergy}. Our numerical study showed that, the energy
eigenfunctions in the linear combination Eq. (\ref{14}) are
orthonormal and $\Psi (a, \tau )$ Eq. (\ref{14}), has a constant norm.
In the linear combination Eq. (\ref{14}), we set $C_n$ equal to one,
for the first eighteen values of $n$ and $C_n$ equal to zero for the
other values of $n$. In Figure \ref{f1}, we show, as an example, the
modulus squared of a wave packet constructed with the first eighteen
energy levels, given in Table \ref{tableenergy}, for $\tau = 1000$,
$L = 6$ and $\Lambda = -0.1$.

\begin{figure}[pb]
\centerline{\psfig{file=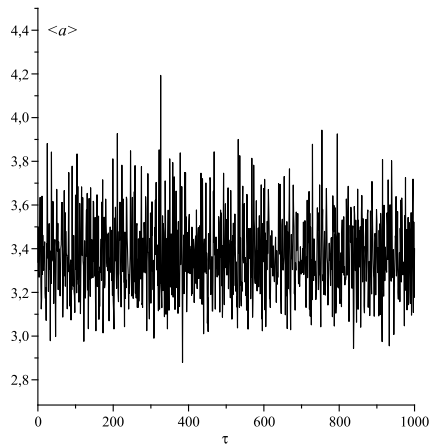,width=6.0cm}}
\vspace*{22pt}
\caption{$\left<a\right>$ computed with the wave packet constructed with the
first eighteen energy levels for the period from $\tau = 0$ until $\tau = 1000$,
$L = 6$ and $\Lambda = -0.1$.
\label{f3}}
\end{figure}

Finally, using the wave packet $\Psi (a, \tau )$ we compute the mean
value for the scale factor $a$, according to the following expression,
\begin{equation}
\left<a\right>(\tau) = \frac{\int_{0}^{\infty}a^{-1}\,|\Psi (a,\tau)|^2 da}
{\int_{0}^{\infty}a^{-2}|\Psi (a,\tau)|^2 da}.
\label{24}
\end{equation}
We computed this quantity for many different time intervals. For all
this different values, we observed that $\left<a\right>$ performs bounded
oscillations and never assume the zero value. Therefore, we confirm
that these models are free from singularities, also, at the quantum level.
As an example, we show in Fig. \ref{f3} the mean value computed with the
wave packet constructed with the first eighteen energy levels, given in
Table \ref{tableenergy}, for the interval from $\tau = 0$ until $\tau = 1000$,
$L = 6$ and $\Lambda = -0.1$.

\section{Conclusions.}
\label{sec:conclusions}

In the present paper, we quantized a Friedmann-Robertson-Walker
model in the presence of a stiff matter perfect fluid and a negative
cosmological constant. We used the variational formalism of Schutz.
The model has spatial sections with negative constant curvature. The
quantization of the model gave rise to a Wheeler-DeWitt equation, for
the scale factor. We found the approximate eigenvalues and eigenfunctions
of that equation by using the Galerkin or spectral method. 
In the present work, we considered only the negative energy eigenvalues 
and their corresponding eigenfunctions. This
choice implies that the energy density of the perfect fluid is negative.
A stiff matter perfect fluid with this property produces a model with a 
bouncing solution, at the classical level, free from an initial 
singularity. After that,
we used the eigenfunctions in order to construct wave packets and
evaluate the time dependent, expected value of the scale factor. We found
that the expected value of the scale factor evolve with bounded oscillations.
Since the expectation value of the scale factor never vanish,
we confirm that this model is free singularities, also, at the quantum level.

\section*{Acknowledgments}

E. V. Corr\^{e}a Silva (Researcher of CNPq, Brazil), G. A. Monerat, G. Oliveira-Neto,
C. Neves and L. G. Ferreira Filho thank CNPq and FAPERJ for partial financial support.
G. Oliveira-Neto thanks FAPEMIG for partial financial support.

\end{document}